\documentstyle[preprint,prb,aps]{revtex}
\begin{document}
%\draft

\title{Interpretation of Nuclear Quadrupole Resonance Spectra in Doped La$_2$CuO$_4$}
 
\author{S. Pliber\v{s}ek and P. F. Meier}
\address{Physics Institute, University of Zurich, CH-8057 Zurich, 
         Switzerland}
%\date{\today}
\maketitle

\begin{abstract}
The nuclear quadrupole resonance (NQR) spectrum of strontium doped La$_2$CuO$_4$ surprisingly
resembles the NQR spectrum of La$_2$CuO$_4$ doped with excess oxygen, both spectra
being dominated by a main peak and one principal satellite peak at similar frequencies.  
Using first-principles cluster 
calculations this is investigated here by calculating the electric 
field gradient (EFG) at the central copper site of the cluster after replacing a lanthanum atom in
the cluster with a strontium atom or adding an interstitial oxygen to the cluster.
In each case the EFG was increased by approximately 10 \% leading unexpectedly to
the explanation that the NQR spectra are only accidentally similar and the
origins are quite different.  Additionally the widths of the peaks in the NQR 
spectra are explained by the different EFG of copper centres remote from the impurity.
A model, based on holes moving rapidly across the planar oxygen atoms, is
proposed to explain the observed increase in frequency of both the main and satellite 
peaks in the NQR spectrum as the doping concentration is increased.
\end{abstract}


%%%%%%%%%%%%%%%%%%%%%%%%%%%%%%%%%%%%%%%%%%%%%%%%%%%%%%%%%%%%%%%%%%%%%%%%%%%%%%%

A large amount of information about static and dynamic properties of 
high-temperature superconducting oxides
has been provided by nuclear
magnetic resonance (NMR) and nuclear quadrupole resonance (NQR) experiments~\cite{review}. 
In certain materials, however, 
some NMR and NQR results are still not completely
understood. In part, this is related to the lack of theoretical support
since, to date, only a few first-principles calculations have addressed the
local charge and spin density distributions.
An example of these unresolved problems is the interpretation of the NQR spectra in doped
La$_2$CuO$_4$. This material becomes superconducting only upon alloying on the
La site or changing the oxygen content. There is therefore considerable interest
in understanding the changes in the local electronic structure that the various
dopants induce.

Experimental observations~\cite{imaisl} of NQR spectra for stoichiometric La$_2$CuO$_4$
show one sharp peak at 33 MHz assigned to $^{63}$Cu. 
Upon doping with
strontium, a second weaker line appears at a somewhat higher 
frequency~\cite{yoshimura,kumagai,nakamura,yoshimura2,song}.
This was called B-line to distinguish it from the main resonance line (A).
Increasing the concentration of dopants causes both peaks to broaden as
well as increase in frequency slightly.  Similar behaviour is observed~\cite{hammel,statt} 
when oxygen is used as the dopant. This produces an additional line which is so 
close in frequency to the B-line frequency
of the Sr-doped spectra  that both lines were attributed to the same physical origin.
Hence it was proposed that holes are localized in CuO$_6$ octahedra
adjacent to the charged dopants. These holes, if pinned at least for times  of
the order of microseconds, would then cause the NQR B-lines. This interpretation was
corroborated by results from ab-initio calculations~\cite{martin} on the magnitude
of the copper NQR frequency which showed that the Sr- and O-dopants themselves had
very different effects on the NQR frequency but the frequency shift for a copper
adjoining a site occupied by a localized hole matched the experimental results 
very closely.
Although there is strong evidence for various kinds of charge inhomogeneities in
doped lanthanum cuprate, the occurrence of static pinned holes proved elusive.
In addition, there are some subtle differences between the NQR spectra of Sr- and O-doped
samples, such as the asymmetry of the electric field gradient of the B-lines, which are
not explained by the above model.

In this work, we report on results of density functional cluster calculations
that address the changes in the electric field gradients (EFGs) in La$_2$CuO$_4$ upon doping.
Four different cluster models were used to simulate various
experimental situations:\\
($M_{\alpha}$) CuO$_6$/Cu$_4$La$_{10}$ represents the La$_2$CuO$_4$ crystal and used to obtain the
EFG at the central copper site\\
($M_{\beta}$) CuO$_6$/Cu$_4$La$_{9}$Sr represents the La$_{2-x}$Sr$_x$CuO$_4$ crystal in the dilute
limit $x \rightarrow 0$\\
($M_{\gamma}$) CuO$_6$/Cu$_4$La$_{10}$ and CuO$_6$/Cu$_4$La$_{9}$Sr, as above, but where positive 
charges are placed on each planar oxygen.  This model is proposed to investigate the influence of
holes that are moving rapidly across the planar oxygen sites\\
($M_{\delta}$) CuO$_7$/Cu$_4$La$_{10}$, the same as CuO$_6$/Cu$_4$La$_{10}$ but where an extra oxygen
atom occupies an interstitial site, representing the La$_2$CuO$_{4+\delta}$ crystal in
the dilute limit $\delta \rightarrow 0$.

Recently, extensive cluster studies for the pure La$_2$CuO$_4$ system have been
presented~\cite{huesserprb}. 
Spin-polarized calculations at the Hartree-Fock (HF) level and with 
the density functional (DF) method with gradient corrections to the exchange and
correlation functionals have
been performed for clusters comprising up to 9 Cu atoms in a CuO$_2$-plane.
It has been shown that an acceptable estimate of the EFG at the copper site
can already be obtained
with DF calculations with a small cluster (CuO$_6$/Cu$_4$La$_{10}$) where a (CuO$_6$)$^{10-}$-ion
is treated with an all-electron basis set and 4 surrounding Cu and 10 La atoms are
represented by bare pseudopotentials. This is the cluster which is chosen as model $M_{\alpha}$
in the present work. It
is embedded in a lattice of 2464 point charges.
The positions of some
point charges were changed in such a way that the correct Madelung
potential in the central region of the cluster was reproduced.
All atomic positions were located according to the tetragonal phase with
lattice constants~\cite{lattice} a = 3.77 {\AA} and c = 13.18 {\AA} and with 
a Cu-O(apex) distance of 2.40 {\AA}.
For the cluster atoms standard triple-zeta (6-311 G) basis sets were employed. 
%For Cu, this
%corresponds to the basis set developed by Wachters~\cite{Wachters}, while
%for O it was given in Ref.~\cite{pople}.
All calculations were spin-polarized and  were performed with the Gaussian94 and Gaussian98 
programs \cite{g98}.
In the DF calculations the correlational functional of Lee, Yang, and Parr~\cite{LYP} was
combined with the Becke~\cite{becke1} exchange functional, which also includes
corrections based on the gradient of the electron density.

For model $M_{\alpha}$
the density functional calculations with the BLYP functional give for $V_{zz}$,
the z-component of the EFG, at the Cu site a value of 1.419 a.u..
This corresponds to a $^{63}$Cu quadrupole frequency of  
35 MHz if a quadrupole moment $Q(^{63}$Cu) = $-.211$ b is adopted~\cite{sternheimer}.
The experimental $^{63}$Cu NMR spectrum of pure La$_2$CuO$_4$ shows~\cite{imaisl}
a narrow peak at 33 MHz.
The good agreement between theory and experiment may be accidental
in view of the small cluster size. Furthermore, there is no general consensus about the precise
value of $Q(^{63}$Cu). Other sources of inaccuracy are due to the limited basis set used
for the representation of the atomic orbitals and the choice of the exchange and correlation
functionals for the treatment of the generalized gradient approximations beyond the
local density approximations.
Since we focus in this paper on the relative
changes of the local electronic structure around the copper atom which are induced by
models for various doping effects, all calculations are carried out at the same level
of approximation.
We expect then that the difference between EFG values calculated for different clusters
will not significantly be influenced by the above inaccuracies.

It should further be remarked that the EFG cannot be explained in a simple one-electron
(or rather one-hole) picture.
The theoretical value results from cancellations between relatively large individual
contributions from the various atomic shells which are listed in the second column of 
Table~\ref{tab:contr}.

It has been shown~\cite{huesserprb} that other choices of the exchange-correlation functional
have little influence on $V_{zz}$. HF calculations, however, produce a $V_{zz}$-value
that is 36 \% higher. This discrepancy between DF and HF results for the EFG at Cu sites
has also been found in previous calculations for the YBa$_2$Cu$_3$O$_7$ system. It is
further discussed in Ref.~\cite{huesser,durban} but is not the topic of the present work which
focusses on the relative changes in $V_{zz}$ upon doping.

To simulate the influence of doping with Sr, the calculations were then repeated for the
(CuO$_6$)$^{10-}$-ion with the La$^{3+}$ on top of the Cu replaced by Sr$^{2+}$ (model $M_{\beta}$).
This simulates a situation where the hole introduced has moved to a place far from the
cluster region or is distributed over several sites.
The resulting changes in the contributions to $V_{zz}$ with respect 
to the undoped system
are given in the third column of Table~\ref{tab:contr}. The EFG is increased by 10.3 \%
which is very close to the observed difference~\cite{} of about 3 MHz in $^{63}\nu_Q$ between the
additional peak (B-line) and the main peak (A-line).
It should be noted that in models $M_{\alpha}$ and $M_{\beta}$ the same atomic positions were used.
Structural changes induced by replacing the lanthanum atom by strontium are thus not
accounted for.

At the Cu sites adjacent to the central Cu ion of the cluster, the replacement of
La$^{3+}$ by Sr$^{2+}$ leads to an increase of $V_{zz}$ by 1 \%. This explains the 
observed broadening of the main peak (A-line) in La$_{2-x}$Sr$_x$CuO$_4$.

Similar cluster models ($M_{\alpha}$ for pure and $M_{\beta}$ for Sr-doped La$_{2}$CuO$_4$) 
have previously been investigated by Martin~\cite{martin} with HF calculations
with results which drastically differ from those of our DF calculations.
We therefore also performed HF calculations for the clusters $M_{\alpha}$ and $M_{\beta}$. 
For the undoped system, our HF result ($V_{zz}$ = 1.94 a.u.) is in broad 
agreement~\cite{rem} with his ($V_{zz}$ = 1.84 a.u.) and also with the value 
reported by Sulaiman et al.~\cite{sulaiman} ($V_{zz}$ = 2.23 a.u.).
For the Sr-doped system, however, the results differ strongly. In Ref.~\cite{martin},
an increase of $V_{zz}$ by 1.5 \% was reported while our HF calculations predict a change
by 5.9 \%. 
%This discrepancy could arise from the different modelling of the
%Sr for which in Ref.~\cite{martin} a 2+ point charge was used while we consistently exchanged the
%pseudopotential for La$^{3+}$ by one for Sr$^{2+}$.

Our DF results thus suggest that the B-line observed in La$_{2}$CuO$_4$ upon doping with
strontium simply originates from the Cu atoms next to the positions of the dopant atoms.
If the models $M_{\alpha}$ and $M_{\beta}$ adequately describe the local electronic structure
around the strontium dopant, a separate oxygen NQR spectrum is to be expected due to
$^{17}$O nuclei
that bridge the strontium atom and its nearest copper atom. For model $M_{\alpha}$, our 
calculations give $V_{zz} = 0.28$ a.u. for the apical oxygens. For model $M_{\beta}$ they
predict $V_{zz} = 0.15$ a.u. for the bridging O(a). 
The number of oxygen nuclei that exhibit this spectrum, however, only amounts to
$p_{\rm{O}} x / 4$ per formula unit where $p_{\rm{O}}$ is the abundance of $^{17}$O nuclei in the sample.
It should be noted, however, that the clusters used in this work were not designed 
for studying the EFGs at the oxygen sites. Indeed, using larger clusters for the 
undoped system, the values of the EFGs at the apical oxygens turn out to 
be smaller and closer to the experimental values of 0.22 a.u.. The pronounced calculated
reduction of $V_{zz}$ 
upon replacing a La by a Sr atom, however, is related to the charge redistribution and will
persist also in more elaborate calculations.

Next we investigate the changes induced by an interstitial oxygen atom O(i) at the
location (a/2,0,c/4). In the tetragonal structure, this site is surrounded by four nearest 
oxygen neighbours at a distance of 2.09 {\AA} for the geometry parameters used 
here, followed by four La 
%at 2.087 {\AA} 
and four Cu.
%at 3.427 {\AA}. 
The latter form two pairs
which lie in adjacent CuO$_2$-planes and have the four O(a) atoms at corresponding apical positions.
This interstitial location for O(i) is undisputed~\cite{tetragonal,chaillout}. It is also evident 
that the squeezing in of the excess oxygen into the interstice will cause a relaxation
of the atoms in the immediate neighborhood. The La positions seem to be unchanged~\cite{chaillout}. 
The EFG value at the copper site
depends crucially on the precise position of the apical oxygen near the interstitial site, even
though the extent of its displacement is disputed~\cite{chaillout,radaelli,cordero}. 
Therefore we determined this position of the O(a)
which also minimised the energy of our cluster model $M_{\delta}$.
For this, 
the cluster CuO$_6$/Cu$_4$La$_{10}$ was extended by including an
additional oxygen O(i).
This model also accounts for that situation where no additional
holes are present in the cluster region. 

The positions of all cluster atoms and point charges of the background were first
frozen except those
of the apical oxygen with coordinates ($x_a,y_a,z_a$) nearest to the O(i) 
which was allowed to move on the plane $y_a = 0$. An energy minimum was found for
$x_a = -0.64$ {\AA} and $z_a = 2.26$ {\AA}. At this position, the distance to the interstitial oxygen is
$2.73$ {\AA}. This corresponds to a tilt of the O(a) by 16$^0$ which is reasonable when compared to
experimental information~\cite{chaillout,radaelli,cordero}. 
Next the positions of the three other apical oxygens coordinated with O(i)
(which are represented by point charges only)
were displaced accordingly. With this geometry, the EFG is no longer axially
symmetric and the eigenvector of the largest component is rotated by 4$^o$ from the z-axis.
The asymmetry of the EFG tensor amounts to $\eta = 0.05$ in approximate agreement with the 
experimental~\cite{statt} value of 0.07.
The largest eigenvalue of the EFG is 1.575~a.u. which is 11.0 \% larger than in model $M_{\alpha}$.
This corresponds to a shift in frequency of 3.8 MHz.
HF calculations by Martin~\cite{martin} which were carried out without taking the local 
distortions around the interstitial oxygens into account, showed an increase of about 5~MHz with
respect to the results for the undoped case.

Our model $M_{\delta}$ can thus account for the occurrence and position of the satellite peak
in the NQR spectrum of La$_2$CuO$_{4+\delta}$. 
%This is again at variance with the HF results of Martin~\cite{martin}. 
The close agreement of the frequencies of the B-lines in strontium- and
oxygen-doped lanthanum cuprate is therefore only accidental. The physical origins 
of the B-lines are quite different in the two cases. This is also reflected in the 
asymmetry parameter $\eta$ which is zero in
model $M_{\beta}$ by symmetry. Repeating the calculations for a cluster with atomic positions 
chosen according to the orthorhombic structure yields an 
asymmetry of $\eta$ = 0.01 for model $M_{\beta}$. 

Since each interstitial oxygen has four nearest copper atoms, the intensity of the B-line in 
La$_2$CuO$_{4+\delta}$ is thus expected to increase with $4 \delta$ which is also in agreement 
with the experiments~\cite{statt}.

Finally we study the shift of the A-line upon doping. 
We adopted model $M_{\gamma}$ where holes rapidly move across the planar oxygen sites. 
For a rough estimation,
positive point charges $x/2$ were placed near the planar oxygen sites in the cluster
CuO$_6$/Cu$_4$La$_{10}$. To be precise, point charges with values x/8 were put at positions where
the 2p$_x$ and 2p$_y$ orbitals of the planar oxygens attain their extremal values.
The calculations were performed for x = 0.01, 0.02, and 0.04, respectively.
In the fourth column of Table~\ref{tab:contr},
the corresponding changes in the contributions to $V_{zz}$ with respect to the undoped system,
which turned out to be linear in x, are listed for x = 0.04.
We obtain a shift
$d (\log V_{zz} ) / dx  \approx 0.7$
%\begin{equation}
%\end{equation}
which is in reasonable agreement with the experimental value
$d (\log \nu_Q) / dx \approx 0.6$ since no account is given to the fact that the lattice
constants also slightly change with the doping concentration.
Applying the same procedure to model $M_{\beta}$ yields $d (\log V_{zz} ) / dx \approx 0.6$.

Our conclusions are as follows. The results obtained with the cluster CuO$_6$/Cu$_4$La$_{9}$Sr
(model $M_{\beta}$) predict an increase of the EFG at the Cu-site immediately adjacent
to the Sr by 10 \%.
This would explain the occurrence and position of the NQR B-line in 
La$_{2-x}$Sr$_x$CuO$_4$ for $x \rightarrow 0$. In addition, the observed broadening
of the A-line with increasing doping concentration could be attributed to
the small shifts calculated for Cu sites not immediately adjacent to the Sr.
%These are our main results. 
Next, the shift of the A-line with increasing $x$ could
be explained by holes moving rapidly across the planar oxygens (model $M_{\gamma}$). 
%At a lower level of confidence,
We have shown that by introducing excess oxygen in La$_2$CuO$_{4+\delta}$ (model $M_{\delta}$), the EFG
at the Cu is also increased by an amount which accidentally is about the same as doping
with Sr produces. Hence our results disagree with previous interpretations of the Cu NQR spectra in 
the doped La$_2$CuO$_{4}$
system, in particular that of Martin~\cite{martin}
which attribute the origin of the B-lines to localized holes that are introduced either by
Sr or O doping.

We express our gratitude to T. A. Claxton for enlightening discussions
and critical reading of the manuscript. We gratefully acknowledge the help of P. H\"usser 
and E. P. Stoll.
This work is partially supported by the Swiss National Science Foundation.

\begin{table}
\caption{Contributions to the EFG component $V_{zz}$ in atomic units
%, calculated with DF/GGA for the cluster CuO$_6$/Cu$_4$La$_{10}$,
from the nuclei within the cluster, from the point charges around the 
cluster, and from the individual shells. The ``remainder''
lists just the (small) rest of all other contributions that cannot be assigned
to a particular shell.
The second column shows the values for pure La$_2$CuO$_4$ (model $M_{\alpha}$). In the third 
and forth column the changes are given which result when one La$^{3+}$ is
replaced by Sr$^{2+}$ (model $M_{\beta}$) and for mobile holes corresponding 
to $x=0.04$ (model $M_{\gamma}$), respectively.}
\begin{tabular}{lrrr}
                             &  pure   &  Sr doped    & holes x=0.04  \\ \hline
Nuclei                       &  0.375     & 0.003     &   0.000   \\
Point charges                &  0.012     & 0.000     & $-0.002$  \\ 
p$_x$, p$_y$                 & $-2.393$   &  0.012    &  0.013    \\
p$_z$                        &   0.770    & 0.028     &  0.003    \\
d$_{x^{2}-y^{2}}$, d$_{xy}$  & $-15.163$  & 0.043     &   0.011   \\
d$_{3z^{2}-r^{2}}$           &    8.829   & 0.040     &   0.011    \\
d$_{yz}$, d$_{zx}$           &    9.003   & 0.002     &   0.001    \\
Remainder                    &  $-0.012 $ & 0.019     &   0.007    \\ \hline
Total                        &    1.419   & 0.147     &   0.044    \\
\end{tabular}
\label{tab:contr}
\end{table}

%%%%%%%%%%%%%%%%%%%%%%%%%%%%%%%%%%%%%%%%%%%%%%%%%%%%%%%%%%%%%%%%%%%%%%%%%%%%%%%


\begin{thebibliography}{99}
\bibitem{rem}
	The value $^{63}\nu_Q = 38.9$ MHz given in Ref.~\cite{martin} was converted into 
	an electric field gradient using  Q($^{63}$Cu) = $-.18$ b for the quadrupole moment.
\bibitem{tetragonal}
        Most of the experimental information is of course available for the orthorhombic
	structure. For comparison, we will use the tetragonal structure.
\end{thebibliography}
\end{document}